# Quantum coherence can be transformed into heat

Xue-Qun Yan[†], Yan-Jiao Du, Wen-Tao Hou, and Xiao-Ming Liu

School of Physical Science and Technology, TianGong University, Tianjin 300387, China

The first law of thermodynamics restates the law of conservation of energy. It partitions the change in energy of a system into two pieces, heat and work. While there is no ambiguity to define heat and work in classical thermodynamics, their classification in the quantum regime is not that obvious. Thus, the first law of thermodynamics becomes problematic in the quantum regime. However, recent studies have shown if contribution of quantum coherence is considered to the change of internal energy of the system, the first law of thermodynamics can be extended to the quantum domain. Here we investigate the new version of first law of thermodynamics for some quantum transformations by using two-level atomic system under non-dissipative channel. In our work we achieve a novel result that quantum coherence can be transformed into heat, and the heat can dissipate into the environments.

## I. INTRODUCTION

When it comes to conservation laws, it is naturally easy to recall the first law of thermodynamics, which is a formulation of the law of energy conversion and conservation. In classical cases, the first law of thermodynamics classifies the changes of energy for macroscopic systems as the work performed by external driving and heat exchanged with the environments [1]. Recently it has been recognized that the law of thermodynamics should be redefined in the quantum cases because the coherence plays a significant and fundamental role [2,3]. It was shown that coherence is independent of work and heat derived from their respective classical analogies, and it is a part of the first law of quantum thermodynamics. Thus, quantum coherence would participate in the conversion of energy in a quantum system. Progress in the past decade has consistently shown that it is possible to construct systems in which thermodynamics coexists with quantum effects [4]. Indeed, coherence is the signature of quantum behavior which is used to drive a wide variety of phenomena and devices. The well-known quantum phenomenon, the wave nature of particles, can be interpreted as manifestations of quantum coherence. More importantly, quantum coherence is regarded as a key ingredient to develop quantum technologies. Coherence is recently considered as an essential resource for quantum process, since it may be consumed to achieve useful tasks [4]. Many novel insights stem from the characterization of quantum coherence as a physical resource. Based on the resource theory of quantum thermodynamics, there have been lots of study works on understanding the roles of coherence in quantum thermodynamics [4-6]; Misra *et al*. reported their study on the role of coherence in quantum thermodynamics [7], in which they analyzed the physical situation that the resource theories of coherence and thermodynamics play completing roles. On the other hand, the effects of coherence on work extraction [8], and in determining the distribution of work done on a quantum system have also been presented [9,10]. There are some studies, in which coherence is treated as a key factor in the operation of quantum thermal machines, such as heat engines and refrigerators [11-15]. It has also been noted that quantum coherence plays important roles in the process of conversion from thermal to electrical power [16]. Moreover, recent works have found that coherence affects the performance of non-adiabatic work protocols [17-20]. Despite many works have demonstrated that quantum coherence can be used as an advantage or a resource for various thermodynamic processes [21-29], the role of quantum coherences in thermodynamics is still not fully understood. Understanding functional role of coherence is an important topic in the field of thermodynamics of quantum systems.

---
[†]E-mail: yanxuequn@tiangong.edu.cn

## II. FIRST LAW OF QUANTUM THERMODYNAMICS

The generalization of thermodynamics to quantum regimes faces challenges ranging from the proper identification of heat and work to the clarification of the role of coherence. How to define work and heat is still a controversial topic in quantum thermodynamics, therefore in quantum regimes it is necessary to revisit these time-honored concepts.

Consider a generic quantum system $S$ with reduced density matrix $\rho_S$ evolving under a Hamiltonian $H_S$ and coupled to an external environment. Since the internal energy of the system can be expressed as $U = \langle \hat{H}_S \rangle = Tr(\hat{\rho}_S \hat{H}_S)$ [30], a common approach is based on the change of the total energy expectation value: $dU = Tr[\hat{\rho}_S d\hat{H}_S + \hat{H}_S d\hat{\rho}_S]$, defining the first term (change of Hamiltonian) as work $dW$ and the second (change of state) as heat $dQ$. This formulation gives an interpretation of the differential form of the first law of thermodynamics. The change of state may be associated with a change of entropy, *i.e.*, heat. However, it is generally believed that the heat defined is not exactly equivalent to classical heat [2]. Because of its non-classical properties, it can be called quantum heat here. Since the coherence plays a unique role in the quantum thermodynamics, when there is quantum coherence in the system both the energetics and the coherence properties must be considered together. So, in quantum thermodynamics, the first law can be redefined as [2]: $dU = dW + dQ + dC$. We can explain this by confirming that quantum coherence, as heat and work, is a form of energy exchanged between system and environment. This relation provides a quantum version of the first law for some specific quantum processes analogous to that of classical thermodynamics. The classical form is: $dU = dW + dQ$, where $dW$ is the amount of work done by external to the system, and $dQ$ is the amount of heat added to the system during an infinitesimal process. Here we should note that $C$ does not has a classical analog, unlike $W$ and $Q$. In contrast to the classical definitions, work and heat are also no longer absolute physical quantities. They depend on the chosen measurement basis as quantum coherence. These quantities can be in principle calculated if the density operator $\hat{\rho}(t)$ and the Hamiltonian $\hat{H}(t)$ of the system are given. Their time dependence in a finite quantum process can be calculated by integration. The expression of change of the internal energy of the system can be written as [2]

$$\Delta U(t) = \sum_n \sum_k \int_0^t \frac{d}{dt'}\left(E_n \rho_k |C_{n,k}|^2\right) dt' \tag{1}$$

where $C_{n,k} = \langle n|k \rangle$. $\{|k\rangle\}$ is the eigenstate basis of the density operator $\hat{\rho}$, and $\rho_k = \langle k|\hat{\rho}|k\rangle$ is the eigenvalues of $\hat{\rho}$, i.e., $\hat{\rho} = \sum_k \rho_k |k\rangle\langle k|$. Here, the Hamiltonian is expressed as $\hat{H}_S = \sum_n E_n |n\rangle\langle n|$, with $E_n = \langle n|\hat{H}_S|n\rangle$ and $|n\rangle$ are the $n$th energy eigenvalue and eigenstate, respectively. The work, heat and quantum coherence in finite-time quantum processes can also be calculated respectively by direct integration [2]:

$$W(t) = \sum_n \sum_k \int_0^t \rho_k |C_{n,k}|^2 \frac{dE_n}{dt'} dt' \tag{2}$$

$$Q(t) = \sum_n \sum_k \int_0^t E_n |C_{n,k}|^2 \frac{d\rho_k}{dt'} dt' \tag{3}$$

and

$$C(t) = \sum_n \sum_k \int_0^t (E_n \rho_k) \frac{d}{dt'} |C_{n,k}|^2 dt' \tag{4}$$

These are the energetic contribution of the dynamics of the work, heat and coherence, respectively. As can be seen, $W(t)$, $Q(t)$ and $C(t)$ depend on the quantity $|C_{n,k}(t)|^2 = |\langle n(t)|k(t)\rangle|^2$. In a quantum process, this quantity varies only if the directions of the basis vectors $|k\rangle$ of the density operator change with respect to the basis vectors $|n\rangle$ of the Hamiltonian.

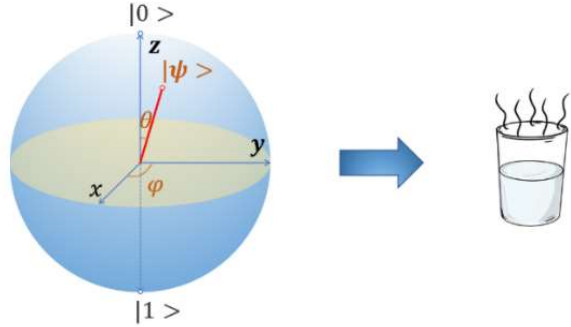

**FIG. 1** (color online). Diagrammatic representation of the coherence transforms into heat. The left diagram represents the Bloch sphere. The qubit $|\psi\rangle$ (coherent superposition state of a two-level quantum system) is represented by a point on the surface of the Bloch sphere, and is defined by a vector having an angle $\theta$ with the polar axis (here $z$) and azimuthal angle φ is on x-y plane. The north and south poles correspond to the pure qubit states $|0\rangle$ and $|1\rangle$, respectively.

As we known that coherence cannot be converted to work through a direct physical process. This phenomenon is known as work locking [31-33]. One may naturally ask whether it can be converted into heat in a direct way, or how it works to the change of the internal energy of the system. Answering this question is the purpose of our work. The basic idea being discussed is shown in Fig. 1. As shown below, we will try to give such a process that they can convert coherence into heat in a direct physical process. To be more concrete, let us consider the case of a qubit under non-dissipative channel first.

## III. PHYSICAL MODEL

Before investigating this problem, we briefly recall the Kraus operator sum representation. Given an initial state for a qubit $\rho(0)$, the evolution under external environments can be expressed in the Kraus operator sum representation [34]: $\varepsilon[\hat{\rho}(0)] = \sum_i \hat{K}_i \hat{\rho}(0) \hat{K}_i^+ = \rho(t)$. The operation elements $\hat{K}_i$ are the Kraus operators associated with the decohering process of a single qubit and satisfy $\sum_i K_i^+ K_i = I$ then $\text{Tr}[\rho(t)] = 1$. If the qubit is under a non-dissipative channel, decoherence occurs in the absence of transfer of energy. More specifically, we first focus on the dynamical evolution of the system under the action of phase damping channel, which the energy eigenstates of the quantum system are invariant with the time evolution, but do accumulate a phase which is proportional to the eigenvalue. For this case, the Kraus operators are given by [34]

$$\hat{K}_1 = \begin{pmatrix} 1 & 0 \\ 0 & \sqrt{1-\gamma} \end{pmatrix}, \quad \hat{K}_2 = \begin{pmatrix} 0 & 0 \\ 0 & \sqrt{\gamma} \end{pmatrix} \tag{5}$$

where $\gamma = 1 - \exp(-\Gamma t)$ with $\Gamma$ denoting a decay rate. If we suppose that in the energy basis $\{|g\rangle, |e\rangle\}$, the initial state is written in the form

$$\hat{\rho}(0) = \begin{pmatrix} \rho_{00} & \rho_{01} \\ \rho_{10} & \rho_{11} \end{pmatrix} \tag{6}$$

then the density matrix as a function of time is given by

$$\hat{\rho}(t) = \begin{pmatrix} \rho_{00} & e^{-\frac{1}{2}\Gamma t} \rho_{01} \\ e^{-\frac{1}{2}\Gamma t} \rho_{10} & \rho_{11} \end{pmatrix} \tag{7}$$

In the following, the system is considered to be a two-level atom, whose ground and excited states, $|g\rangle$ and $|e\rangle$, have energies $E_g$ and $E_e$, respectively, so that the Hamiltonian is given by $\hat{H}_S = E_g |g\rangle\langle g| + E_e |e\rangle\langle e|$. We consider the case that the atom is assumed to be initially prepared in the pure state $|\psi(0)\rangle = \cos\theta |g\rangle + \sin\theta |e\rangle$. Since the energy eigenvalues are constant ($E_n = E_g, or E_e$), no work is done by means of Eq. (2), i.e., $W = 0$. And the phase-damping channel indues a loss of quantum coherence without net energy exchange between the system and environment, thus in this process the internal energy of the system remains unchanged, that is, $\Delta U = 0$. In terms of the redefined first law of thermodynamics, we have that $dQ = -dC$. It shows that the coherence can be completely converted into the heat of the system in this process. In a finite quantum process, by calculating the eigenvalues and eigenstates of $\hat{\rho}(t)$, we can obtain the analytical expressions for $C$ and $Q$ by means of Eqs. (3) and (4) (see Appendix A Information for details). For an illustration, some parameters are fixed. We have set the initial state with $\theta = \pi/6$. The results are presented in Fig. 2. The Fig. 2 shows that since the internal energy is unchanged, in phase damping channel the heat of atomic absorption is always equal to the extracted coherence. This fact can indicate, in our opinion, the coherence can be used as a resource to generate heat.

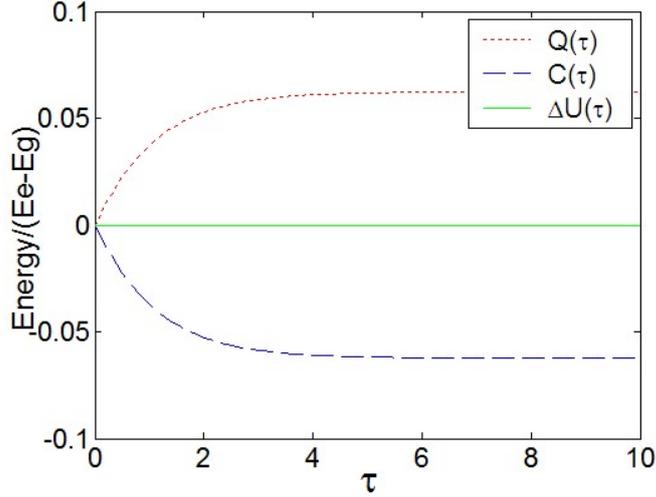

**FIG. 2** (color online). Phase damping process. The heat, work and internal energy, as function of the dimensionless time $\tau \equiv \Gamma t$ in the case of $\theta = \pi/6$. The dotted (online red) curve corresponds to the heat exchanged between the atom system and the environment, Q, the dashed (online blue) curve corresponds to the quantum coherence of the system, while the solid (online green) line corresponds to the internal energy of the system $\Delta U$. Following the classical definition, Q is negative for heat leaves the system, so if heat added to the system, Q is positive. The process causes the coherence to be extracted and converted into heat of the system.

In order to further investigate this physical insight, we also consider the case of a two-level atom under Pauli maps (phase flip, bit flip, and bit-phase flip channels) [34]. These channels are non-dissipative channels, thus there is not energy exchange between the system and environment in these processes. The Kraus operators $\hat{K}_i$ for phase flip, bit flip, and bit-phase flip channels are given by Table 1. For the channels, the explicit time dependence of the dephasing factor is $p = 1 - \exp(-\Gamma t)$ with $\Gamma$ being the dephasing rate. Next, we consider only phase flip channel as noise model, since the evolutions of the state $\rho(t)$ under bit flip and bit-phase flip channel are symmetric with that of phase flip channel.

| Channel | Kraus operators | |
|---|---|---|
| Phase flip | $K_1 = \sqrt{1-p}I$ | $K_2 = \sqrt{p}\sigma_z$ |
| Bit flip | $K_1 = \sqrt{1-p}I$ | $K_2 = \sqrt{p}\sigma_x$ |
| Bit-phase flip | $K_1 = \sqrt{1-p}I$ | $K_2 = \sqrt{p}\sigma_y$ |

Table 1. Kraus operators $\hat{K}_i$ for phase flip, bit flip, and bit-phase flip channels in terms of $p$. Here, $I$ is unit operator, and $\sigma_x$, $\sigma_y$ and $\sigma_z$ are Pauli operators respectively.

The density operator as a function of time, in the energy basis $\{|g\rangle, |e\rangle\}$ under phase flip channel, has the following forms

$$\rho(t) = \begin{pmatrix} \rho_{00} & (2e^{-\Gamma t} - 1)\rho_{01} \\ (2e^{-\Gamma t} - 1)\rho_{10} & \rho_{11} \end{pmatrix} \tag{8}$$

As in the forward case, we consider that the atom is assumed to be initially prepared in the state $|\psi(0)\rangle = \cos\theta|g\rangle + \sin\theta|e\rangle$. Similarly, if we calculate the eigenvalues and eigenstates of $\hat{\rho}(t)$ in (8), then we can use Eqs. (3) and (4) again to obtain $C$ and $Q$ (see Appendix B Information for details). Fig. 3 shows the results of $C$ and $Q$ for $\theta = \pi/6$. As can be seen from the figure that in phase flip channel the heat of atomic absorption is still always equal to the extracted coherence despite there is a decay over time. It is clear that the results are qualitatively the same as seen in Fig. 2.

These qualitative and numerical analyses lead us to a bold conclusion that the quantum coherence of an atom can be converted into heat transferred into the environment.

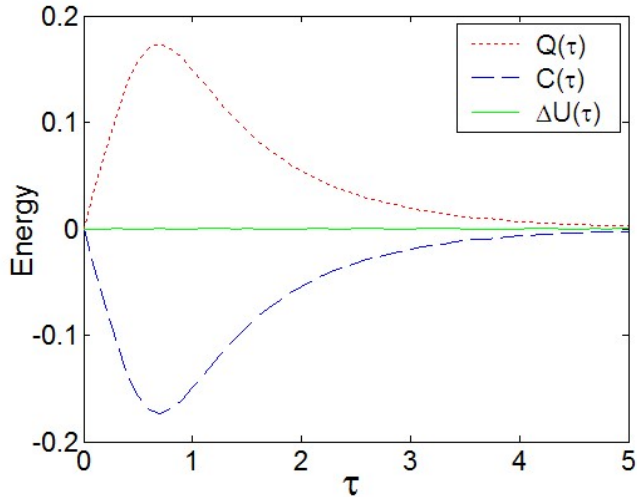

**FIG. 3** (color online). Same as Fig. 2 but for phase flip channel. For specific, we set $E_g = 0$, $E_e = 1$.

## IV. CONCLUSIONS

We have investigated the first law of quantum thermodynamics for some quantum transformation in the framework of two-level atomic systems under non-dissipative channel. We can clearly see that due to the presence of coherences, the first law of thermodynamics is found to be unsatisfied in the quantum regime and must be corrected by introduced the coherence instead. It is well known that heat is not the energy that an object contains, but rather refers to the amount of energy transferred from one object to another. In the classical domain, energy is transferred by heat and work. However, in the quantum regime, coherence may also be involved in energy conversion. Although the quantum heat is different from classical heat, we believe that in some cases, the quantum heat must contain some characteristics of classical heat. We therefore speculate from our analysis that under certain circumstances, such as the non-dissipative channel, coherence can be converted into the heat, which dissipates into the environment. It is fair to say, however the fundamental meaning of quantum coherence and heat for our understanding of the first law of quantum thermodynamics is still awaiting discovery. While there is no doubt that our results give a new insight into this problem.

Finally, we briefly mention that physically, phase damping can be employed to describe, for example, what happens when a photon scatters randomly as it travels through a waveguide, or how electronic states in an atom are perturbed upon interacting with distant electrical charges. Thus, this suggests that our theoretical results would be realized through the possible experiments.

**ACKNOWLEDGEMENTS**

We thank Tiangong University 2017 degree and graduate education reform project, Project No. Y20170702.

## APPENDIX A: THERMODYNAMICS UNDER PHASE DAMPING CHANNEL

Now we calculate in detail the evolution of heat and quantum coherence of a two-level atom under the action of phase damping channel. As pointed out in the main text, the time evolution of the atom is given by (7). In order to evaluate $Q(t)$ and $C(t)$, we need calculate the eigenvalues of $\hat{\rho}(t)$, which can be found as follows

$$\rho_0(t) = \frac{1}{2}\left(\rho_{00} + \rho_{11} + \sqrt{m}\right) \tag{A1}$$

and

$$\rho_1(t) = \frac{1}{2}\left(\rho_{00} + \rho_{11} - \sqrt{m}\right) \tag{A2}$$

as well as the respective eigenvectors

$$|k_0(t)\rangle = \frac{1}{\sqrt{\left(\rho_{11}-\rho_{00}-\sqrt{m}\right)^2 + 4e^{-\Gamma t}\rho_{10}^2}}\left[\left(\rho_{00} - \rho_{11} + \sqrt{m}\right)|g\rangle + 2e^{-\frac{1}{2}\Gamma t}\rho_{10}|e\rangle\right] \tag{A3}$$

and

$$|k_1(t)\rangle = \frac{1}{\sqrt{\left(\rho_{11}-\rho_{00}+\sqrt{m}\right)^2 + 4e^{-\Gamma t}\rho_{10}^2}}\left[\left(\rho_{00} - \rho_{11} - \sqrt{m}\right)|g\rangle + 2e^{-\frac{1}{2}\Gamma t}\rho_{10}|e\rangle\right] \tag{A4}$$

where $m = \rho_{00}^2 - 2\rho_{00}\rho_{11} + \rho_{11}^2 + 4e^{-\Gamma t}\rho_{01}\rho_{10}$.

Since the initial state is $|\psi(0)\rangle = \cos\theta|g\rangle + \sin\theta|e\rangle$, using the results of Eqs. (A1) to (A4), we can calculate the heat exchanged between the atom and the environment as a function of dimensionless scaled time $\tau \equiv \Gamma t$ by means of Eq. (3) (here we take $\theta = \pi/6$),

$$Q(\tau) = E_g\left[\int_0^\tau |\langle g|k_0(\tau')\rangle|^2 \frac{d}{d\tau'}\rho_0(\tau')\,d\tau' + \int_0^\tau |\langle g|k_1(\tau')\rangle|^2 \frac{d}{d\tau'}\rho_1(\tau')\,d\tau'\right]$$

$$+ E_e\left[\int_0^\tau |\langle e|k_0(\tau')\rangle|^2 \frac{d}{d\tau'}\rho_0(\tau')\,d\tau' + \int_0^\tau |\langle e|k_1(\tau')\rangle|^2 \frac{d}{d\tau'}\rho_1(\tau')\,d\tau'\right]$$

$$= \frac{(E_e - E_g)}{8}\left[\tau + \log 4 - \log(3 + e^\tau)\right] \tag{A5}$$

The result is shown in Fig. 2. In what follows, we calculate the energetic contribution of the dynamics of coherence in this example. From Eq. (4) we obtain that

$$C(\tau) = E_g\left[\int_0^\tau \rho_0(\tau')\frac{d}{d\tau'}|\langle g|k_0(\tau')\rangle|^2\,d\tau' + \int_0^\tau \rho_1(\tau')\frac{d}{d\tau'}|\langle g|k_1(\tau')\rangle|^2\,d\tau'\right]$$

$$+E_e\left[\int_0^\tau \rho_0(\tau')\frac{d}{d\tau'}|\langle e|k_0(\tau')\rangle|^2\,d\tau' + \int_0^\tau \rho_1(\tau')\frac{d}{d\tau'}|\langle e|k_1(\tau')\rangle|^2\,d\tau'\right]$$

$$= \frac{(E_e - E_g)}{8}\left[-\tau - \log 4 + \log(3 + e^\tau)\right] \tag{A6}$$

The result is also shown in Fig. 2 along with that for the internal energy.

### APPENDIX B: THERMODYNAMICS UNDER PHASE FLIP CHANNEL

Next let us calculate the evolution of heat and quantum coherence of a two-level atom under the action of phase flip channel. The eigenvalues of the equation (8), $\hat{\rho}(t)$, can be found to be

$$\rho_0(t) = \frac{1}{2}\left(\rho_{00} + \rho_{11} + \sqrt{m}\right) \tag{B1}$$

and

$$\rho_1(t) = \frac{1}{2}\left(\rho_{00} + \rho_{11} - \sqrt{m}\right) \tag{B2}$$

as well as the respective eigenvectors

$$|k_0(t)\rangle = \frac{1}{\sqrt{(\rho_{11}-\rho_{00}-\sqrt{m})^2 + 4\rho_{10}^2(2e^{-\Gamma t}-1)^2}}\left[(\rho_{00} - \rho_{11} + \sqrt{m})|g\rangle + 2(2e^{-\Gamma t} - 1)^2\rho_{10}|e\rangle\right] \tag{B3}$$

and

$$|k_1(t)\rangle = \frac{1}{\sqrt{(\rho_{11}-\rho_{00}+\sqrt{m})^2 + 4\rho_{10}^2(2e^{-\Gamma t}-1)^2}}\left[(\rho_{00} - \rho_{11} - \sqrt{m})|g\rangle + 2(2e^{-\Gamma t} - 1)^2\rho_{10}|e\rangle\right] \tag{B4}$$

where $m = \rho_{00}^2 - 2\rho_{00}\rho_{11} + \rho_{11}^2 + 4(2e^{-\Gamma} - 1)^2\rho_{01}\rho_{10}$.

In the followings, we can calculate the heat exchanged between the atom and the environment as a function of dimensionless scaled time $\tau \equiv \Gamma t$ by means of Eq. (3). The initial state is $|\psi(0)\rangle = \cos\theta|g\rangle + \sin\theta|e\rangle$ (here we take $\theta = \pi/6$), then the evolution of heat is given

$$Q(\tau) = \frac{E_g}{16}\left[-4 + 4e^{-\tau}\sqrt{e^{2\tau} - 3e^\tau + 3} - 2\tau + \log(e^{2\tau} - 3e^\tau + 3)\right]$$

$$+\frac{E_g}{16}\left[4 - 4e^{-\tau}\sqrt{e^{2\tau} - 3e^\tau + 3} - 2\tau + \log(e^{2\tau} - 3e^\tau + 3)\right]$$

$$+\frac{E_e}{16}\left[-4 + 4e^{-\tau}\sqrt{e^{2\tau} - 3e^\tau + 3} - \log(3e^{-2\tau} - 3e^{-\tau} + 1)\right]$$

$$+\frac{E_e}{16}\left[4 - 4e^{-\tau}\sqrt{e^{2\tau} - 3e^\tau + 3} - \log(3e^{-2\tau} - 3e^{-\tau} + 1)\right] \tag{B5}$$

The dynamics of coherence can be obtained by

$$C(\tau) = \frac{E_g}{16}\left[1 + 2\tau - \log(3 - 3e^\tau + e^{2\tau}) - \frac{e^\tau}{\sqrt{e^{2\tau} - 3e^\tau + 3}}\right]$$

$$+\frac{E_g}{16}\left[-1 + 2\tau - \log(3 - 3e^\tau + e^{2\tau}) + \frac{e^\tau}{\sqrt{e^{2\tau} - 3e^\tau + 3}}\right]$$

$$+\frac{E_e}{16}\left[-1 - 2\tau + \log(3 - 3e^\tau + e^{2\tau}) + \frac{e^\tau}{\sqrt{e^{2\tau} - 3e^\tau + 3}}\right]$$

$$+\frac{E_e}{16}\left[1-2\tau+\log(3-3e^\tau+e^{2\tau})-\frac{e^\tau}{\sqrt{e^{2\tau}-3e^\tau+3}}\right] \quad (B6)$$

For specific, if we set $E_g = 0$, $E_e = 1$, then

$$Q(\tau) = \frac{1}{8}[-\log(1+3e^{-2\tau}-3e^{-\tau})] \quad (B7)$$

and

$$C(\tau) = \frac{1}{8}[\log(3+e^{2\tau}-3e^\tau)-2\tau] \quad (B8)$$

These results are plotted in Fig. 3 along with that for the internal energy.

---

[1] M. Plischke and B. Bergersen, *Equilibrium Statistical Physics* (World Scientific Publishing Co. te. Ltd.，2003).
[2] B. L. Bernardo, Unravelling the role of coherence in the first law of quantum thermodynamics, Phys. Rev. E **102**, 062152 (2020).
[3] B. L. Bernardo, Relating heat and entanglement in strong-coupling thermodynamics, Phys. Rev. E **104**, 044111 (2021).
[4] A. Streltsov, G. Adesso, and M. B. Plenio, Colloquium: Quantum coherence as a resource, Rev. Mod. Phys. **89**, 041003 (2017).
[5] J. Goold, M. Huber, A. Riera, L. del Rio, and P. Skrzypczyk, The role of quantum information in thermodynamics-a topical review, J. Phys. A **49**, 143001 (2016).
[6] G. Gour, M. P. Müller, V. Narasimhachar, R. W. Spekkens, and N. Y. Halpern, The resource theory of informational nonequilibrium in thermodynamics, Phys. Rep. **583**, 1 (2015).
[7] A. Misra, U. Singh, S. Bhattacharya, and A. K. Pati, Energy cost of creating quantum coherence, Phys. Rev. A **93**, 052335 (2016).
[8] P. Kammerlander and J. Anders, Coherence and measurement in quantum thermodynamics, Sci. Rep. **6**, 22174 (2016).
[9] P. Solinas and S. Gasparinetti, Full distribution of work done on a quantum system for arbitrary initial states, Phys. Rev. E **92**, 042150 (2015).
[10] P. Solinas and S. Gasparinetti, Probing quantum interference effects in the work distribution, Phys. Rev. A **94**, 052103 (2016).
[11] M. O. Scully, K. R. Chapin, K. E. Dorfman, M. B. Kim, and A. Svidzinsky, Quantum heat engine power can be increased by noise-induced coherence, Proc. Natl. Acad. Sci. U.S.A. **108**, 15097 (2011).
[12] S. Rahav, U. Harbola, and S. Mukamel, Heat fluctuations and coherences in a quantum heat engine, Phys. Rev. A **86**, 043843 (2012).
[13] J. Um, K. E. Dorfman, and H. Park, Coherence-enhanced quantum-dot heat engine, Phys. Rev. Research **4**, L032034 (2022).
[14] P. Bayona-Pena, Thermodynamics of a continuous quantum heat engine: Interplay between population and coherence, Phys. Rev. A **104**, 042203 (2021).
[15] H. Tajima and K. Funo, Superconducting-like Heat Current: Effective Cancellation of Current-Dissipation Trade-Off by Quantum Coherence, Phys. Rev. Lett. **127**, 190604 (2021).

[16] O. Karlström, H. Linke, G. Karlström, and A. Wacker, Increasing thermoelectric performance using coherent transport, Phys. Rev. B **84**, 113415 (2011).

[17] G. Francica, F. C. Binder, G. Guarnieri, M. T. Mitchison, J. Goold, and F. Plastina, Quantum coherence and ergotropy, Phys. Rev. Lett. **125**, 180603 (2020).

[18] F. H. Kamin, F. T. Tabesh, S. Salimi, and A. C. Santos, Entanglement, coherence and charging process of quantum batteries, Phys. Rev. E **102**, 052109 (2020).

[19] H. Kwon, H. Jeong, D. Jennings, B. Yadin, and M. S. Kim, Clock–Work Trade-Off Relation for Coherence in Quantum Thermodynamics, Phys. Rev. Lett. **120**, 150602 (2018).

[20] M. Lostaglio, An introductory review of the resource theory approach to thermodynamics, Rep. Prog. Phys. **82**, 114001 (2019).

[21] A. Streltsov, G. Adesso, and M. B. Plenio, Colloquium: Quantum coherence as a resource, Rev. Mod. Phys. **89**, 041003 (2017).

[22] M. O. Scully, M. S. Zubairy, G. S. Agarwal, and H. Walther, Extracting work from a single heat bath via vanishing quantum coherence, Science **299**, 862 (2003).

[23] J. Åberg, Catalytic Coherence, Phys. Rev. Lett. **113**, 150402 (2014).

[24] R. Uzdin, Coherence-Induced Reversibility and Collective Operation of Quantum Heat Machines Via Coherence Recycling, Phys. Rev. Appl. **6**, 024004 (2016).

[25] K. Korzekwa, M. Lostaglio, J. Oppenheim, and D. Jennings, The extraction of work from quantum coherence, New J. Phys. **18**, 023045 (2016).

[26] T. Van Vu and K. Saito, Thermodynamics of Precision in Markovian Open Quantum Dynamics, Phys. Rev. Lett. **128**, 140602 (2022).

[27] M. Perarnau-Llobet, K. V. Hovhannisyan, M. Huber, P. Skrzypczyk, N. Brunner, and A. Acín, Extractable Work from Correlations, Phys. Rev. X **5**, 041011 (2015).

[28] A. Streltsov, H. Kampermann, S. Wölk, M. Gessner, and D. Bruß, Maximal coherence and the resource theory of purity, New J. Phys. **20**, 053058 (2018).

[29] B. Çakmak, A. Manatuly, and Ö. E. Müstecaplıoglu, Thermal production, protection, and heat exchange of quantum coherences, Phys. Rev. A **96**, 032117 (2017).

[30] R. Alicki, The quantum open system as a model of the heat engine, J. Phys. A: Math. Gen. **12**, L103 (1979).

[31] M. Horodecki and J. Oppenheim, Fundamental limitations for quantum and nanoscale thermodynamics, Nat. Commun. **4**, 2059 (2013).

[32] M. Lostaglio, D. Jennings, and T. Rudolph, Description of quantum coherence in thermodynamic processes requires constraints beyond free energy, Nat. Commun. **6**, 6383 (2015).

[33] P. Skrzypczyk, A. J. Short, and S. Popescu, Work extraction and thermodynamics for individual quantum systems, Nat. Commun. **5**, 4185 (2014).

[34] M. A. Nielsen and I. L. Chuang, *Quantum computation and quantum information* (Cambridge University Press 2000).